%% file: Allertonv3.tex
\documentclass[10pt,conference]{IEEEtran}

\usepackage{epsf,psfrag,amssymb,amsfonts,cite,amsmath,eufrak,yfonts}
\input macro
\def\scalefig#1{\epsfxsize #1\textwidth}
\allowdisplaybreaks[4]
\begin{document}
\title{On the Capacity of MIMO Interference Channels}

\author{\authorblockN{Xiaohu Shang}
\authorblockA{
Department of EECS\\
Syracuse University\\
 Email: xshang@syr.edu} \and
 \authorblockN{Biao Chen}
\authorblockA{
Department of EECS\\
Syracuse University\\
 Email: bichen@syr.edu}\and
\authorblockN{Gerhard Kramer}
\authorblockA{Bell Laboratories\\
Alcatel-Lucent\\
 Email: gkr@research.bell-labs.com}
\and
\authorblockN{H. Vincent Poor}
\authorblockA{
Department of EE\\
Princeton University\\
Email: poor@princeton.edu}}\maketitle{\footnotetext{This research
was supported in part by the National Science Foundation under
Grants CCF-05-46491, ECS-05-01534, ANI-08-38807 and
CNS-06-25637.}}


\begin{abstract}
The capacity region of a multiple-input-multiple-output
interference channel (MIMO IC) in which the channel matrices are
square and invertible is studied. The capacity region for
 strong interference is established where the definition of strong interference
 parallels that of
scalar channels. Moreover, the sum-rate capacity for Z
interference, noisy interference, and mixed interference is
established. These results generalize known results for the scalar
Gaussian IC.
\end{abstract}

\section{Introduction}

The interference channel (IC) models the situation in which
transmitters communicate with their respective receivers while
generating interference to all other receivers. This channel model
was mentioned in \cite[Section 14]{Shannon:61Berkeley} and its
capacity region is still generally unknown.

In \cite{Carleial:75IT} Carleial showed that interference does not
reduce capacity when it is very strong. This result follows
because the interference can be decoded and subtracted at each
receiver before decoding the desired message. Later Han and
Kobayashi \cite{Han&Kobayashi:81IT} and Sato \cite{Sato:81IT}
showed that the capacity region of the strong interference channel
is the same as the capacity region of a compound multiple access
channel. In both above cases, the interference is fully decoded at
both receivers.

When the interference is not strong, the capacity region is
unknown. The best inner bound is by Han and Kobayashi
\cite{Han&Kobayashi:81IT}, which was later simplified by Chong
\etal in \cite{Chong-etal:08IT,Kramer:06Zurich}. Etkin \etal and
Telatar and Tse showed that Han and Kobayashi's inner bound is
within one bit of the capacity region of scalar Gaussian ICs
\cite{Etkin-etal:07IT_submission,Telatar&Tse:07ISIT}. Various
outer bounds have been developed in
\cite{Kramer:04IT,Etkin-etal:07IT_submission,Telatar&Tse:07ISIT,Shang-etal:07IT_submission,
Motahari&Khandani:08IT_submission,Annapureddy&Veeravalli:08IT_submission}.

Special ICs such as the degraded IC and the ZIC have been studied
in \cite{Sato:78IT,Costa:85IT}. The sum-rate capacity for the ZIC
was established  in \cite{Sato:78IT,Sason:04IT}, and Costa proved
the equivalence of the ZIC and the degraded IC for the scalar
Gaussian case \cite{Costa:85IT}. A recent result in
\cite{Shang-etal:07IT_submission,
Motahari&Khandani:08IT_submission,Annapureddy&Veeravalli:08IT_submission}
has shown that if a simple condition is satisfied, then treating
interference as noise can achieve the sum-rate capacity.
\cite{Motahari&Khandani:08IT_submission} and
\cite{Weng&Tuninetti:08ITA} derived the sum-rate capacity for
mixed interference, i.e., one receiver experiences strong
interference and the other experiences weak interference.

\begin{figure}[h] \centerline{
\begin{psfrags}
\psfrag{x1}[c]{$\xp_1$} \psfrag{x2}[c]{$\xp_2$}
\psfrag{y1}[c]{$\yp_1$} \psfrag{y2}[c]{$\yp_2$}
\psfrag{n1}[c]{$\zp_1$} \psfrag{n2}[c]{$\zp_2$} \psfrag{+}[c]{$+$}
\psfrag{g11}[c]{$\Hbf_1$} \psfrag{g12}[c]{$\Hbf_3$}
\psfrag{g21}[c]{$\Hbf_2$} \psfrag{g22}[c]{$\Hbf_4$}
\scalefig{.35}\epsfbox{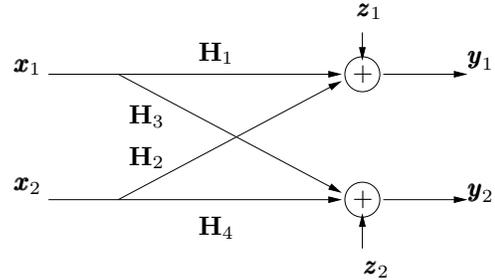}
\end{psfrags}}
\caption{\label{fig:model} The MIMO IC.}
\end{figure}

In this paper, we study the sum-rate capacity of the two-user
Gaussian multiple-input-multiple-output (MIMO) IC shown in Fig.
\ref{fig:model}. The received signals are defined as
 \bqa
&&\hspace{-.2in}\yp_1=\Hbf_1\xp_1+\Hbf_2\xp_2+\zp_1\nn\\
&&\hspace{-.2in}\yp_2=\Hbf_3\xp_1+\Hbf_4\xp_4+\zp_2,\label{eq:model}\eqa
where $\xp_i,i=1,2,$ is the transmitted signal of user $i$ which
is subject to an average block power constraint $P_i$;
$\zp_i,i=1,2$ is a Gaussian random vector with zero mean and
identity covariance matrix; and $\Hbf_{j}, j=1,\dots,4,$ are the
channel matrices. For simplicity, we assume that the $\Hbf_j$'s
are real and that $\Hbf_1$ and $\Hbf_4$ are invertible. However we
remark that one can generalize our results to non-invertible or
rectangular channel matrices (see Remark 1).

For the MIMO IC Telatar and Tse \cite{Telatar&Tse:07ISIT} showed
that Han and Kobayashi's region is within one bit per receive
antenna of the capacity region. Some upper bounds were discussed
in \cite{Vishwanath&Jafar:04ITW} and some lower bounds on the
sum-rate capacity based on Han and Kobayashi's region were
discussed in \cite{Shang-etal:06IT}. However capacity results for
the MIMO IC are still lacking. In our work, assuming the channel
matrices are invertible, we derive the sum-rate capacity with
noisy-interference, strong interference and mixed interference, as
well as one-sided interference. The capacity region of the MIMO IC
with strong interference is also obtained.

The rest of the paper is organized as follows: we present our main
results and proofs in Section II and III; numerical results are
given in Section IV, and we conclude in Section V.

Before proceeding we introduce some notation which will be used in
the paper. \bi \item Italic font $X$ denotes a scalar; and the
bold fonts $\xp$ and $\Xbf$ denote vectors and matrices
respectively. \item $\Abf\succeq\Bbf$ means that $\Abf-\Bbf$ is
positive semi-definite. \item $\Ibf$ denotes the identity matrix
and $\0bf$ denotes the zero matrix. \item $|\Xbf|$, $\Xbf^T$,
$\Xbf^H$, $\Xbf^{-1}$, $\Xbf^{-T}$ denote respectively the
determinant, transpose, conjugate transpose, inverse, and
transpose inverse of the matrix $\Xbf$. \item
$\xp^n=\left[\xp_1^T,\xp_2^T,\dots,\xp_n^T\right]^T$ is a long
vector which consists of a sequence of vectors $\xp_i, i=1,\dots,
n$.\item $||\Smat||$ denotes the size of the set $\Smat$. \item
$abs(\cdot)$ denotes the absolute value. \item
$\xp\sim\Nmat\left(\0bf,\Sigmabf\right)$ means that the random
vector $\xp$ is Gaussian distributed with zero mean and covariance
matrix $\Sigmabf$. \item $E[\cdot]$ denotes expectation;
$\textrm{Cov}(\cdot)$ denotes covariance matrix; $I(\cdot;\cdot)$
denotes mutual information; $h(\cdot)$ denotes differential
entropy with the logarithm base $e$ and
$\log(\cdot)=\log_e(\cdot)$.

\ei

\section{Main Results}

\begin{theorem}
For the MIMO IC defined in (\ref{eq:model}), and where the channel
matrices $\Hbf_1$ and $\Hbf_4$ are square and invertible, the
sum-rate capacity is achieved by treating interference as noise at
both receivers if for any covariance matrices $\Sbf_i,i=1,2$, with
$\tr(\Sbf_i)\leq P_i$, the following conditions are satisfied:
\bqa
&&\hspace{-.4in}\max_{\alphabf^H\alphabf=1}abs\left(\alphabf^H\Mbf^{-\frac{1}{2}}\Wbf_1\Mbf^{-\frac{1}{2}}\alphabf\right)\leq\frac{1}{2},\label{eq:condition1}\eqa
and\bqa
&&\hspace{-.4in}\max_{\alphabf^H\alphabf=1}abs\left(\alphabf^H\Mbf^{-\frac{1}{2}}\Wbf_2\Mbf^{-\frac{1}{2}}\alphabf\right)\leq\frac{1}{2},\label{eq:condition2}\eqa
where \bqa &&\hspace{-.3in}
\Mbf=\Ibf-\Abf_1\Abf_1^T-\Abf_2\Abf_2^T,\label{eq:M}\\
&&\hspace{-.3in}\Wbf_1=\Abf_1^T\Abf_2^T,\label{eq:W1}\\
&&\hspace{-.3in}\Wbf_2=\Abf_2^T\Abf_1^T,\\
&&\hspace{-.3in}\Abf_1=\left(\Ibf+\Hbf_2\Sbf_2\Hbf_2^T\right)\Hbf_1^{-T}\Hbf_3^T\quad\textrm{and}\\
&&\hspace{-.3in}\Abf_2=\left(\Ibf+\Hbf_3\Sbf_1\Hbf_3^T\right)\Hbf_4^{-T}\Hbf_2^T.\eqa
The sum-rate capacity is the solution of the following
optimization problem \bqa
\max &&\frac{1}{2}\log\left|\Ibf+\Hbf_1\Sbf_1\Hbf_1^T\left(\Ibf+\Hbf_2\Sbf_2\Hbf_2^T\right)^{-1}\right|\nn\\
&&\hspace{.2in}+\frac{1}{2}\log\left|\Ibf+\Hbf_4\Sbf_2\Hbf_4^T\left(\Ibf+\Hbf_3\Sbf_1\Hbf_3^T\right)^{-1}\right|\nn\\
\textrm{subject to} && \textrm{tr}(\Sbf_1)\leq
P_1,\textrm{tr}(\Sbf_2)\leq P_2,\nn\\
&&\Sbf_1\succeq \0bf, \Sbf_2\succeq\0bf.\label{eq:opt}\eqa
\label{theorem:NIsum}
\end{theorem}

{\em Remark 1:} 
Theorem \ref{theorem:NIsum} can be generalized to the MIMO ICs
with the channel matrices $\Hbf_1$ and $\Hbf_4$ being
non-invertible or rectangular. In those cases, two additional
conditions must be satisfied such that the matrices $\Abf_1$ and
$\Abf_2$ exist. This result will be reported in a subsequent
paper.

{\em Remark 2:} In the scalar case, if we have $\Hbf_1=\Hbf_4=1$,
$\Hbf_2=\sqrt{a}$, $\Hbf_3=\sqrt{b}$, from (\ref{eq:condition1})
and (\ref{eq:condition2}) we obtain  \bqa
\sqrt{a}(1+bP_1)+\sqrt{b}(1+aP_2)\leq 1.\label{eq:scalar}\eqa
Therefore Theorem \ref{theorem:NIsum} is an extension of the
noisy-interference sum-rate capacity of the scalar IC
\cite{Shang-etal:07IT_submission,Annapureddy&Veeravalli:08IT_submission,
Motahari&Khandani:08IT_submission} to the MIMO IC.

{\em Remark 3:} Theorem \ref{theorem:NIsum} is also valid by
replacing the power constraint with the covariance matrix
constraint. This extension applies to all the following theorems.

\begin{theorem}
For the MIMO IC defined in (\ref{eq:model}), and where the channel
matrices $\Hbf_1$ and $\Hbf_4$ are square and invertible, the
sum-rate capacity is achieved by treating interference as noise at
both receivers, if for any covariance matrices $\Sbf_i,i=1,2$,
with $\tr(\Sbf_i)\leq P_i$, there exist symmetric positive
definite matrices $\Sigmabf_1$ and $\Sigmabf_2$ satisfying the
following conditions \bqa &&
\hspace{-.4in}\Sigmabf_{1}\preceq\Ibf-\Abf_{2}\Sigmabf_{2}^{-1}\Abf_{2}^T\quad\textrm{and}\label{eq:condition2_1}\\
&&\hspace{-.4in}\Sigmabf_2\preceq\Ibf-\Abf_1\Sigmabf_1^{-1}\Abf_1^T,\label{eq:condition2_2}\eqa
where $\Abf_1$ and $\Abf_2$ are defined in Theorem
\ref{theorem:NIsum}. \label{theorem:NIsum2}
\end{theorem}

Theorem \ref{theorem:NIsum2} is another description of a
sufficient condition for single-user detection to be sum-rate
optimal. It can be shown that for the scalar case,
(\ref{eq:condition2_1}) and (\ref{eq:condition2_2}) reduce to
(\ref{eq:scalar}).

\begin{theorem}
For the MIMO IC defined in (\ref{eq:model}) with $\Hbf_3=\0bf$ and
$\Hbf_2$ and $\Hbf_4$ square and invertible, the sum-rate capacity
is \bqa C^*=\max_{\substack{\textrm{tr}(\Sbf_1)\leq
P_1\\\textrm{tr}(\Sbf_2)\leq
P_2}}\frac{1}{2}\log\left|\Ibf+\Hbf_1\Sbf_1\Hbf_1^T\left(\Ibf+\Hbf_2\Sbf_2\Hbf_2^T\right)^{-1}\right|\nn\\
+\frac{1}{2}\log\left|\Ibf+\Hbf_4\Sbf_2\Hbf_4^T\right|,\eqa if the
following condition is satisfied \bqa
\Hbf_2^T\Hbf_2\prec\Hbf_4^T\Hbf_4.\eqa Furthermore, \bqa
&&\hspace{-.5in}C^*=\max_{\textrm{tr}\left(\Sbf_1\right)\leq
P_1,\textrm{tr}\left(\Sbf_2\right)\leq
P_2}\nn\\
&&\hspace{-.42in}\min\left\{\begin{array}{c}
  \dfrac{1}{2}\log\left|\Ibf+\Hbf_1\Sbf_1\Hbf_1^T+\Hbf_2\Sbf_2\Hbf_2^T\right|, \\
  \dfrac{1}{2}\log\left|\Ibf+\Hbf_1\Sbf_1\Hbf_1^T\right|+\dfrac{1}{2}\log\left|\Ibf+\Hbf_4\Sbf_2\Hbf_4^T\right|\\
\end{array}\right\}\eqa if \bqa
\Hbf_2^T\Hbf_2\succeq\Hbf_4^T\Hbf_4.\eqa \label{theorem:ZGIC}
\end{theorem}

Theorem \ref{theorem:ZGIC} gives the sum-rate capacity of a MIMO
ZIC. Specifically, when $\Hbf_2^T\Hbf_2\prec\Hbf_4^T\Hbf_4$ we
consider the interference to be weak and the sum-rate capacity can
be achieved by treating the interference as noise. When
$\Hbf_2^T\Hbf_2\succeq\Hbf_4^T\Hbf_4$ we consider the interference
to be strong and the sum-rate capacity can be achieved by fully
decoding the interference.

\begin{theorem}
For the MIMO IC defined in (\ref{eq:model}), and where the channel
matrices are square and invertible, if
$\Hbf_2^T\Hbf_2\succeq\Hbf_4^T\Hbf_4$ and
$\Hbf_3^T\Hbf_3\succeq\Hbf_1^T\Hbf_1$, then the sum-rate capacity
is
 \bqa
&&\hspace{-.5in}C^*=\max_{\textrm{tr}\left(\Sbf_1\right)\leq
P_1,\textrm{tr}\left(\Sbf_2\right)\leq
P_2}\nn\\
&&\hspace{-.42in}\min\left\{\begin{array}{c}
  \dfrac{1}{2}\log\left|\Ibf+\Hbf_1\Sbf_1\Hbf_1^T+\Hbf_2\Sbf_2\Hbf_2^T\right|, \\
  \dfrac{1}{2}\log\left|\Ibf+\Hbf_3\Sbf_1\Hbf_3^T+\Hbf_4\Sbf_2\Hbf_4^T\right|, \\
  \dfrac{1}{2}\log\left|\Ibf+\Hbf_1\Sbf_1\Hbf_1^T\right|+\dfrac{1}{2}\log\left|\Ibf+\Hbf_4\Sbf_2\Hbf_4^T\right|\\
\end{array}\right\}.\eqa
\label{theorem:GICstrong}
\end{theorem}

Theorem \ref{theorem:GICstrong} shows that if
$\Hbf_2^T\Hbf_2\succeq\Hbf_4^T\Hbf_4$ and
$\Hbf_3^T\Hbf_3\succeq\Hbf_1^T\Hbf_1$ is satisfied, then the
receivers experience strong interference. Thus the channel acts as
a compound MIMO multiple access channel and the sum-rate capacity
is achieved by fully decoding the interference at both users.

\begin{theorem}
For the MIMO IC defined in (\ref{eq:model}), and where the channel
matrices are square and invertible, if
$\Hbf_2^T\Hbf_2\prec\Hbf_4^T\Hbf_4$ and
$\Hbf_3^T\Hbf_3\succeq\Hbf_1^T\Hbf_1$, then the sum-rate capacity
is \bqa
&&\hspace{-.5in}C^*=\max_{\textrm{tr}\left(\Sbf_1\right)\leq
P_1,\textrm{tr}\left(\Sbf_2\right)\leq
P_2}\nn\\
&&\hspace{-.42in}\min\left\{\begin{array}{c}
\dfrac{1}{2}\log\left|\Ibf+\Hbf_3\Sbf_1\Hbf_3^T+\Hbf_4\Sbf_2\Hbf_4^T\right|, \\
  \hspace{-.1in}\dfrac{1}{2}\log\left|\Ibf+\Hbf_1\Sbf_1\Hbf_1^T\left(\Ibf+\Hbf_2\Sbf_2\Hbf_2^T\right)^{-1}\right| \\
  \hspace{1.0in}+\dfrac{1}{2}\log\left|\Ibf+\Hbf_4\Sbf_2\Hbf_4^T\right|\\
  \end{array}\right\}.\eqa
\label{theorem:GICmixed}
\end{theorem}

Theorem \ref{theorem:GICmixed} gives the sum-rate capacity of the
MIMO IC with mixed interference
$\Hbf_2^T\Hbf_2\prec\Hbf_4^T\Hbf_4$ and
$\Hbf_3^T\Hbf_3\succeq\Hbf_1^T\Hbf_1$. The sum-rate capacity is
achieved by treating interference as noise at the receiver that
experiences weak interference and fully decoding the interference
at the receiver that experiences strong interference.

\begin{theorem}
For the MIMO IC defined in (\ref{eq:model}) with $\Hbf_3=0$ and
all other channel matrices being square and invertible, if
$\Hbf_2^T\Hbf_2\succeq\Hbf_4^T\Hbf_4$, then the capacity region is
\bqa \hspace{-.0in}\bigcup_{\substack{\tr\left(\Sbf_1\right)\leq
P_1\\\tr\left(\Sbf_2\right)\leq P_2}} \left\{\begin{array}{c}
  R_1\leq\dfrac{1}{2}\log\left|\Ibf+\Hbf_1\Sbf_1\Hbf_1^T\right| \\
  R_2\leq\dfrac{1}{2}\log\left|\Ibf+\Hbf_4\Sbf_2\Hbf_4^T\right| \\
  R_1+R_2\leq\dfrac{1}{2}\log\left|\Ibf+\Hbf_1\Sbf_1\Hbf_1^T+\Hbf_2\Sbf_2\Hbf_2^T\right| \\
\end{array}\right\}\nn\eqa

\label{theorem:ZICstrongRegion}

\end{theorem}

\begin{theorem}
For the MIMO IC defined in (\ref{eq:model}), and where the channel
matrices $\Hbf_2$ and $\Hbf_4$ are square and invertible, if
$\Hbf_2^T\Hbf_2\succeq\Hbf_4^T\Hbf_4$ and
$\Hbf_3^T\Hbf_3\succeq\Hbf_1^T\Hbf_1$, the capacity region is \bqa
\bigcup_{\substack{\tr\left(\Sbf_1\right)\leq
P_1\\\tr\left(\Sbf_2\right)\leq P_2}}\left\{\begin{array}{c}
  R_1\leq\dfrac{1}{2}\log\left|\Ibf+\Hbf_1\Sbf_1\Hbf_1^T\right| \\
  R_2\leq\dfrac{1}{2}\log\left|\Ibf+\Hbf_4\Sbf_2\Hbf_4^T\right| \\
  R_1+R_2\leq\dfrac{1}{2}\log\left|\Ibf+\Hbf_1\Sbf_1\Hbf_1^T+\Hbf_2\Sbf_2\Hbf_2^T\right| \\
  R_1+R_2\leq\dfrac{1}{2}\log\left|\Ibf+\Hbf_3\Sbf_1\Hbf_3^T+\Hbf_4\Sbf_2\Hbf_4^T\right| \\
\end{array}\right\}\nn\eqa

\label{theorem:ICstrongRegion}

\end{theorem}

Theorems \ref{theorem:ZICstrongRegion} and
\ref{theorem:ICstrongRegion} give the capacity region of the MIMO
ZIC and MIMO IC under strong interference.

Finally we connect the MIMO IC with the parallel Gaussian
interference channel (PGIC), which is a special case of
(\ref{eq:model}) with all $\Hbf_i$'s being diagonal matrices. In
\cite{Shang-etal:08Globecom} we present conditions for which
single user detection for each sub-channel is sum-rate optimal
under the assumption that the coding and decoding is independent
across sub-channels. The following theorem proves that independent
coding and decoding is indeed sum-rate optimal under
noisy-interference if some conditions are satisfied.

\begin{theorem}
For the MIMO IC defined in (\ref{eq:model}) with
$\Hbf_i=diag\left(h_{i1},\dots,h_{it}\right), i=1,\dots,4$, let
$P_{1i}^*$ and $P_{2i}^*$ be the optimal solution of the following
optimization problem \bqa \max &&
\frac{1}{2}\sum_{i=1}^t\left[\log\left(1+\frac{h_{1i}^2P_{1i}}{1+h_{2i}^2P_{2i}}\right)
\right.\nn\\
&&\hspace{.6in}\left.+\log\left(1+\frac{h_{4i}^2P_{2i}}{1+h_{3i}^2P_{1i}}\right)\right]\nn\\
\textrm{subject to}&& \sum_{1=1}^tP_{1i}\leq P_1,\quad
\sum_{1=1}^tP_{2i}\leq P_2\nn\\
&& P_{1i}\geq 0,\quad P_{2i}\geq 0.\eqa Then the sum-rate capacity
is \bqa
&&\hspace{-.3in}C^*=\sum_{i=1}^tC_i(P_{1i}^*,P_{2i}^*)\\
&&\hspace{-.1in}=\frac{1}{2}\sum_{i=1}^t\left[\log\left(1+\frac{h_{1i}^2P_{1i}^*}{1+h_{2i}^2P_{2i}^*}\right)
+\log\left(1+\frac{h_{4i}^2P_{2i}^*}{1+h_{3i}^2P_{1i}^*}\right)\right]\nn\eqa
if \bqa &&\hspace{-.1in}
abs\left(h_{1i}h_{2i}\right)\left(1+h_{3i}^2P_{1i}^*\right)+abs\left(h_{3i}h_{4i}\right)\left(1+h_{2i}^2P_{2i}^*\right)\nn\\
&&\hspace{1in}\leq
abs\left(h_{1i}h_{4i}\right),\label{eq:PGICsepCondi}\eqa   and
\bqa \bigcap_{i=1}^t\partial C_i(P_{1i}^*,P_{2i}^*)\neq\phi, \eqa
for all $i=1,\dots,t$, where $\partial C_i(P_{1i}^*,P_{2i}^*)$
denotes the subdifferential of $C_i(\cdot,\cdot)$ at point
$(P_{1i}^*,P_{2i}^*)$, and $\phi$ denotes the empty set. The
notion of subdifferential follows that in
\cite{Rockafellar:book}.\label{theorem:PGIC}
\end{theorem}

 Theorem
\ref{theorem:PGIC} illustrates that if each sub-channel (each
antenna pair in MIMO IC) satisfies the noisy-interference
condition, then independent decoding at each sub-channel  with
single-user detection achieves the sum-rate capacity. Theorem
\ref{theorem:PGIC} shows the conditions for {\em independent}
coding and single-user detection across sub-channels to be
optimal.


\section{Proof of the Main Results}
\subsection{Preliminaries}
We introduce some lemmas that we use to prove our main results.



\begin{lemma}\cite[Lemma 1]{Thomas:87IT} Let $\xp_1,\dots,\xp_n$ be zero-mean random vectors and denote the
covariance matrix of the stacked vector
$\left[\xp_1^T,\dots,\xp_n^T\right]^T$ as $\Kbf$. Let $\Smat$ be a
subset of $\{1,2,\dots,n\}$ and $\bar\Smat$ be its complement.
Then we have\bqa
h\left(\xp_{\Smat}\left|\xp_{\bar\Smat}\right.\right)\leq
h\left(\xp^*_{\Smat}\left|\xp^*_{\bar\Smat}\right.\right),\eqa
where $\left[\xp_1^T,\dots,\xp_n^T\right]^T\sim
\Nmat\left(\0bf,\Kbf\right)$.\label{lemma:conditional2}
\end{lemma}

\begin{lemma}
Let
$\xp_i^n=\left[\xp_{i,1}^T,\dots,\xp_{i,n}^T\right]^T,i=1,\dots,k$,
be $k$ stacked random vectors each of which consists of $n$
vectors. Let $\yp^{n}=\left[\yp_1^T,\dots,\yp_n^T\right]^T$ be $n$
Gaussian random vectors with covariance matrix\bqa
\sum_{i=1}^k\lambda_i\Cov\left(\xp^n_i\right)=\Cov\left(\yp^{n}\right),\label{eq:cvxCov}\eqa
where $\sum_{i=1}^k\lambda_i=1,\lambda_i\geq 0$. Let $\Smat$ be a
subset of $\{1,2,\dots,n\}$ and $\bar\Smat$ be its complement.
Then we have\bqa \sum_{i=1}^k\lambda_i
h\left(\xp_{i,\Smat}\left|\xp_{i,\bar\Smat}\right.\right)\leq
h\left(\yp_\Smat\left|\yp_{\bar\Smat}\right.\right).\eqa
\label{lemma:generalconcave}
\end{lemma}

The proof of Lemma \ref{lemma:generalconcave} is given in the
appendix. Lemma \ref{lemma:generalconcave} shows the concave-like
property of the conditional entropy
$h\left(\xp_\Smat\left|\xp_{\bar\Smat}\right.\right)$ over the
covariance matrix $\Cov\left(\xp^n\right)$.

Consider a special case of Lemma \ref{lemma:generalconcave} with
$n=1$, $\bar\Smat$ being the empty set and $\lambda_i=1/k$. We
obtain the following lemma.

\begin{lemma} Let $\xp^k$ be a set of $k$ random vectors. Then
\bqa h\left(\xp^k\right)\leq k\cdot
h\left(\widehat\xp^*\right),\eqa where $\widehat\xp^*$ is a
Gaussian vector with the covariance matrix \bqa
\textrm{Cov}\left(\widehat\xp^*\right)=\frac{1}{k}\sum_{i=1}^k\textrm{Cov}\left(\xp_i\right).\eqa
\label{lemma:noconditonal}
\end{lemma}

Let $n=2,||\Smat||=||\bar\Smat||=1$ and $\lambda_i=1/k$. We obtain
another special case of Lemma \ref{lemma:generalconcave}.

\begin{lemma}
Let $\xp^k$ and $\yp^k$ be two sequences of random vectors. Then
we have
 \bqa h\left(\yp^k\left|\xp^k\right.\right)\leq
k\cdot h\left(\widehat\yp^*\left|\widehat\xp^*\right.\right),\eqa
where $\widehat\xp^*$ and $\widehat\yp^*$ are Gaussian vectors
with the joint covariance matrix \bqa
\textrm{Cov}\left[\begin{array}{c}
  \widehat\xp^* \\
  \widehat\yp^* \\
\end{array}\right]=\frac{1}{k}\sum_{i=1}^k\textrm{Cov}\left[\begin{array}{c}
  \xp_i \\
  \yp_i \\
\end{array}\right].\eqa\label{lemma:conditionaldirect}
\end{lemma}

The proof is straightforward from Lemma \ref{lemma:generalconcave}
by noticing that
$h\left(\yp^k\left|\xp^k\right.\right)\leq\sum_{i=1}^kh\left(\yp_i\left|\xp_i\right.\right)$.

\begin{lemma}\cite[Lemma II.2]{Diggavi&Cover:01IT} Let
$\xp^*\sim\Nmat\left(\0bf,\Kbf_x\right)$, and let $\zp$ and
$\zp^*$ be real random vectors (independent of $\xp^*$) with the
same covariance matrix $\Kbf_z$. If
$\zp^*\sim\Nmat\left(\0bf,\Kbf_z\right)$, and $\zp$ has any other
distribution with covariance matrix $\Kbf_z$ then \bqa
I\left(\xp^*;\xp^*+\zp\right)\geq
I\left(\xp^*;\xp^*+\zp^*\right).\eqa If $\Kbf_z\succ\0bf$, then
equality is achieved if and only if
$\zp\sim\Nmat\left(\0bf,\Kbf_z\right)$. \label{lemma:worstnoise}
\end{lemma}

\begin{lemma}
Let $\xp^n$ be a sequence of $n$ zero mean random vectors. Let
$\zp$ and $\tilde\zp$ be two independent Gaussian random vectors
and $\zp^n$ and $\tilde\zp^n$ be two sequences of random vectors
each independent and identically distributed (i.i.d.) as $\zp$ and
$\tilde\zp$ respectively, then \bqa
&&\hspace{-.3in}h\left(\xp^n+\zp^n\right)-h\left(\xp^n+\zp^n+\tilde\zp^n\right)\nn\\
&&\hspace{.2in} \leq
nh\left(\widehat\xp^*+\zp\right)-nh\left(\widehat\xp^*+\zp+\tilde\zp\right),\eqa
where $\widehat\xp^*$ is a zero mean Gaussian random vector with
covariance matrix \bqa
\Cov\left(\widehat\xp^*\right)=\frac{1}{n}\sum_{i=1}^n\Cov\left(\xp_i\right).\eqa\label{lemma:opt}
\end{lemma}
The proof is given in the Appendix.

\begin{lemma} $%
\left[\begin{array}{ll}
  \Ibf & \Abf \\
  \Abf^T & \Bbf \\
\end{array}%
\right]\succeq \0bf$ if and only if $\Bbf\succeq\Abf^T\Abf$.
\label{lemma:pd}
\end{lemma}\vspace{.1in}
The proof is omitted.

\begin{lemma}\cite[Theorem 5.2]{Engwerda-etal:93LA&A}
Suppose $\Wbf$ is nonsingular and $\Mbf$ is positive definite.
Then the matrix equation \bqa \Xbf+\Wbf^H\Xbf^{-1}\Wbf=\Mbf \eqa
has a positive definite solution $\Xbf$ if and only if \bqa
\max_{\alphabf^H\alphabf=1}abs\left(\alphabf^H\Mbf^{-\frac{1}{2}}\Wbf\Mbf^{-\frac{1}{2}}\alphabf\right)\leq\frac{1}{2}.\eqa\label{lemma:pdsolution}
\end{lemma}

\subsection{Proof of Theorem \ref{theorem:NIsum}}
Suppose the channel is used $n$ times. The transmit and receive
vector sequences are denoted by $\xp_i^n$ and $\yp_i^n$ for user
$i$, $i=1,2$. For the $j$th use of the channel, the covariance
matrix of $\xp_{i,j}$ is denoted as $\Sbf_{i,j}$, $j=1,\dots,n$,
and we use the power constraints \bqa
\sum_{j=1}^n\textrm{tr}\left(\Sbf_{i,j}\right)\leq nP_i.\eqa

From Fano's inequality we have that the achievable sum rate
$R_1+R_2$ must satisfy \bqa &&\hspace{-.2in}n(R_1+R_2)-n\epsilon\nn\\
&&\hspace{-.2in}\leq
I\left(\xp_1^n;\yp_1^n\right)+I\left(\xp_2^n;\yp_2^n\right)\nn\\
&&\hspace{-.2in}\leq
I\left(\xp_1^n;\yp_1^n,\Hbf_3\xp_1^n+\np_1^n\right)+I\left(\xp_2^n;\yp_2^n,\Hbf_2\xp_2^n+\np_2^n\right)\nn\\
&&\hspace{-.2in}=h\left(\Hbf_3\xp_1^n+\np_1^n\right)-h(\np_1^n)+h\left(\yp_1^n\left|\Hbf_3\xp_1^n+\np_1^n\right.\right)\nn\\
&&-h\left(\Hbf_2\xp_2^n+\zp_1^n\left|\np_1^n\right.\right)+h\left(\Hbf_2\xp_2^n+\np_2^n\right)-h(\np_2^n)\nn\\
&&+h\left(\yp_2^n\left|\Hbf_2\xp_2^n+\np_2^n\right.\right)-h\left(\Hbf_3\xp_1^n+\zp_2^n\left|\np_2^n\right.\right)\label{eq:original}
\eqa where
$\zp_i^n=\left[\zp_{i,1}^T,\zp_{i,2}^T,\dots,\zp_{i,n}^T\right]^T,
i=1,2$, with all the $\zp_{i,j}, j=1,\dots,n$ independent of each
other.
$\np_i^n=\left[\np_{i,1}^T,\np_{i,2}^T,\dots,\np_{i,n}^T\right]$,
and $\np_{i,j}$ are i.i.d. Gaussian vectors with zero mean and
covariance matrices $\Sigmabf_{i}$. We further let $\np_{i}$ to be
correlated with $\zp_{i}$, and
$E\left[\zp_i\np_{i}^T\right]=\Abf_{i}$. We can write the joint
distribution of $\zp_{i}$ and $\np_{i}$ as \bqa
\left[\begin{array}{c}
  \zp_{i} \\
  \np_{i} \\
\end{array}\right]\sim \Nmat\left(\0bf,\left[\begin{array}{cc}
  \Ibf & \Abf_{i} \\
  \Abf_{i}^T & \Sigmabf_{i} \\
\end{array}\right]\right),\qquad i=1,2,\label{eq:jointZN}\eqa
and we have\bqa
\textrm{Cov}\left(\zp_i\left|\np_{i}\right.\right)=\Ibf-\Abf_{i}\Sigmabf_{i}^{-1}\Abf_{i}^T.\eqa
Let \bqa
\Sigmabf_{1}=\Ibf-\Abf_{2}\Sigmabf_{2}^{-1}\Abf_{2}^T;\label{eq:equation1}
\eqa so we have \bqa
\textrm{Cov}\left(\np_{1}\right)=\textrm{Cov}\left(\zp_{2}\left|\np_{2}\right.\right).\eqa
Since $\np_{1,j}$ is independent of $\np_{1,k}$ and $\zp_{2,j}$ is
independent of $\np_{2,k}$ for any $j\neq k$, we have \bqa
\textrm{Cov}\left(\np_{1}^n\right)=\textrm{Cov}\left(\zp_{2}^n\left|\np_{2}^n\right.\right).\label{eq:eqncovariance}\eqa
Therefore we have \bqa
&&\hspace{-.2in}h\left(\Hbf_3\xp_1^n+\np_1^n\right)-h\left(\Hbf_3\xp_1^n+\zp_2^n\left|\np_2^n\right.\right)=0.\label{eq:difference1}\eqa
Similarly, let \bqa
\Sigmabf_{2}=\Ibf-\Abf_{1}\Sigmabf_{1}^{-1}\Abf_{1}^T;\label{eq:equation2}\eqa
so we have \bqa
&&h\left(\Hbf_4\xp_2^n+\np_2^n\right)-h\left(\Hbf_2\xp_2^n+\zp_1^n\left|\np_1^n\right.\right)=0.\label{eq:difference2}\eqa
Therefore if (\ref{eq:equation1}) and (\ref{eq:equation2}) hold,
(\ref{eq:difference1}) and (\ref{eq:difference2}) are constants
regardless of the distribution of $\xp_1^n$ and $\xp_2^n$. Then we
can write \bqa
&&\hspace{-.4in}h\left(\Hbf_3\xp_1^n+\np_1^n\right)-h\left(\Hbf_3\xp_1^n+\zp_2^n\left|\np_2^n\right.\right)\nn\\
&&\hspace{.2in}=nh\left(\Hbf_3\widehat\xp_1^*+\np_1\right)-nh\left(\Hbf_3\widehat\xp_1^*+\zp_2\left|\np_2\right.\right)\label{eq:constdiff1}\\
&&\hspace{-.4in}h\left(\Hbf_2\xp_2^n+\np_2^n\right)-h\left(\Hbf_2\xp_2^n+\zp_1^n\left|\np_1^n\right.\right)\nn\\
&&\hspace{.2in}=nh\left(\Hbf_2\widehat\xp_2^*+\np_2\right)-nh\left(\Hbf_2\widehat\xp_2^*+\zp_1\left|\np_1\right.\right),\label{eq:constdiff2}\eqa
where $\widehat\xp_1^*$ and $\widehat\xp_2^*$ are zero mean
Gaussian vectors with respective covariance matrices \bqa
\textrm{Cov}\left(\widehat\xp_1^*\right)=\frac{1}{n}\sum_{i=1}^n\Sbf_{1,i}\triangleq\widehat\Sbf_1^*,\eqa
and\bqa
\textrm{Cov}\left(\widehat\xp_2^*\right)=\frac{1}{n}\sum_{i=1}^n\Sbf_{2,i}\triangleq\widehat\Sbf_2^*.\eqa

Next by Lemma \ref{lemma:conditionaldirect} we have \bqa
&&\hspace{-.3in}h\left(\yp_1^n\left|\Hbf_3\xp_1^n+\np_1^n\right.\right)\nn\\
&&\hspace{-.3in}=h\left(\Hbf\xp_1^n+\Hbf_2\xp_2^n+\zp_1^n\left|\Hbf_3\xp_1^n+\np_1^n\right.\right)\nn\\
&&\hspace{-.3in}\leq nh\left(\Hbf_1\widehat\xp_{1}^*+\Hbf_2\widehat\xp_{2}^*+\zp_1\left|\Hbf_3\widehat\xp_{1}^*+\np_1\right.\right)\nn\\
&&\hspace{-.3in}=\frac{n}{2}\log\left|\Hbf_1\widehat\Sbf_{1}^*\Hbf_1^T+\Hbf_2\widehat\Sbf_{2}^*\Hbf_2^T+\Ibf-\left(\Hbf_1\widehat\Sbf_{1}^*\Hbf_3^T+\Abf_1\right)\right.\nn\\
&&\hspace{-.25in}\left.\cdot\left(\Hbf_3\widehat\Sbf_{1}^*\Hbf_3^T+\Sigmabf_1\right)^{-1}\left(\Hbf_3\widehat\Sbf_{1}^*\Hbf_1^T+\Abf_1^T\right)\right|+\frac{n}{2}\log2\pi.\label{eq:conden1}
\eqa Similarly, we obtain \bqa
&&\hspace{-.3in}h\left(\yp_2^n\left|\Hbf_2\xp_2^n+\np_2^n\right.\right)\nn\\
&&\hspace{-.3in}\leq \frac{n}{2}\log\left|\Hbf_4\widehat\Sbf_{2}^*\Hbf_4^T+\Hbf_3\widehat\Sbf_{1}^*\Hbf_3^T+\Ibf-\left(\Hbf_4\widehat\Sbf_{2}^*\Hbf_2^T+\Abf_2\right)\right.\nn\\
&&\hspace{-.25in}\cdot\left.\left(\Hbf_2\widehat\Sbf_{2}^*\Hbf_2^T+\Sigmabf_2\right)^{-1}\left(\Hbf_2\widehat\Sbf_{2}^*\Hbf_4^T+\Abf_2^T\right)\right|+\frac{n}{2}\log2\pi.\label{eq:conden2}
\eqa On substituting (\ref{eq:constdiff1})-(\ref{eq:conden2}) into
(\ref{eq:original}) we have \bqa &&\hspace{-.3in} R_1+R_2-\epsilon\nn\\
&&\hspace{-.3in}\leq\frac{1}{2}\log\left|\Hbf_3\widehat\Sbf_1^*\Hbf_3^T+\Sigmabf_1\right|-\frac{1}{2}\log\left|\Sigmabf_1\right|\nn\\
&&\hspace{-.2in}-\frac{1}{2}\log\left|\Hbf_2\widehat\Sbf_2^*\Hbf_2^T+\Ibf-\Abf_1\Sigmabf_1^{-1}\Abf_1^T\right|\nn\\
&&\hspace{-.2in}+\frac{1}{2}\log\left|\Hbf_1\widehat\Sbf_{1}^*\Hbf_1^T+\Hbf_2\widehat\Sbf_{2}^*\Hbf_2^T+\Ibf-\left(\Hbf_1\widehat\Sbf_{1}^*\Hbf_3^T+\Abf_1\right)\right.\nn\\
&&\hspace{-.2in}\left.\cdot\left(\Hbf_3\widehat\Sbf_{1}^*\Hbf_3^T+\Sigmabf_1\right)^{-1}\left(\Hbf_3\widehat\Sbf_{1}^*\Hbf_1^T+\Abf_1^T\right)\right|\nn\\
&&\hspace{-.2in}+\frac{1}{2}\log\left|\Hbf_2\widehat\Sbf_2^*\Hbf_2^T+\Sigmabf_2\right|-\frac{1}{2}\log\left|\Sigmabf_2\right|\nn\\
&&\hspace{-.2in}-\frac{1}{2}\log\left|\Hbf_3\widehat\Sbf_1^*\Hbf_3^T+\Ibf-\Abf_2\Sigmabf_2^{-1}\Abf_2^T\right|\nn\\
&&\hspace{-.2in}+\frac{1}{2}\log\left|\Hbf_4\widehat\Sbf_{2}^*\Hbf_4^T+\Hbf_3\widehat\Sbf_{1}^*\Hbf_3^T+\Ibf-\left(\Hbf_4\widehat\Sbf_{2}^*\Hbf_2^T+\Abf_2\right)\right.\nn\\
&&\hspace{-.2in}\cdot\left.\left(\Hbf_2\widehat\Sbf_{2}^*\Hbf_2^T+\Sigmabf_2\right)^{-1}\left(\Hbf_2\widehat\Sbf_{2}^*\Hbf_4^T+\Abf_2^T\right)\right|\nn\\
&&\hspace{-.3in}\stackrel{(a)}=\frac{1}{2}\log\left|\Hbf_3\widehat\Sbf_1^*\Hbf_3^T+\Sigmabf_1\right|-\frac{1}{2}\log\left|\Sigmabf_1\right|\nn\\
&&\hspace{-.2in}-\frac{1}{2}\log\left|\Ibf+\Hbf_2\widehat\Sbf_2^*\Hbf_2^T\right|-\frac{1}{2}\log\left|\Ibf-\Hbf_1^{-T}\Hbf_3^T\Sigmabf_1^{-1}\Abf_1^{T}\right|\nn\\
&&\hspace{-.2in}+\frac{1}{2}\log\left|\Ibf+\Hbf_1\widehat\Sbf_{1}^*\Hbf_1^T+\Hbf_2\widehat\Sbf_2^*\Hbf_2^T\right|+\frac{1}{2}\log\left|\Ibf\right.\nn\\
&&\hspace{-.2in}\left.-\Hbf_1^{-T}\Hbf_3^T\left(\Hbf_3\widehat\Sbf_{1}^*\Hbf_3^T+\Sigmabf_1^T\right)^{-1}\left(\Hbf_3\widehat\Sbf_{1}^*\Hbf_1^T+\Abf_1^T\right)\right|\nn\\
&&\hspace{-.2in}+\cdots\nn\\
&&\hspace{-.3in}\stackrel{(b)}=\frac{1}{2}\log\left|\Hbf_3\widehat\Sbf_1^*\Hbf_3^T+\Sigmabf_1\right|-\frac{1}{2}\log\left|\Sigmabf_1\right|\nn\\
&&\hspace{-.2in}-\frac{1}{2}\log\left|\Ibf+\Hbf_2\widehat\Sbf_2^*\Hbf_2\right|-\frac{1}{2}\log\left|\Ibf-\Sigmabf_1^{-1}\Abf_1^{T}\Hbf_1^{-T}\Hbf_3^T\right|\nn\\
&&\hspace{-.2in}+\frac{1}{2}\log\left|\Ibf+\Hbf_1\widehat\Sbf_{1}^*\Hbf_1^T+\Hbf_2\widehat\Sbf_2^*\Hbf_2^T\right|+\frac{1}{2}\log\left|\Ibf\right.\nn\\
&&\hspace{-.2in}\left.-\left(\Hbf_3\widehat\Sbf_{1}^*\Hbf_3^T+\Sigmabf_1^T\right)^{-1}\left(\Hbf_3\widehat\Sbf_{1}^*\Hbf_1^T+\Abf_1^T\right)\Hbf_1^{-T}\Hbf_3^T\right|\nn\\
&&\hspace{-.2in}+\cdots\nn\\
&&\hspace{-.2in}=\frac{1}{2}\log\left|\Ibf+\Hbf_1\widehat\Sbf_{1}^*\Hbf_1^T\left(\Ibf+\Hbf_2\widehat\Sbf_2^*\Hbf_2^T\right)^{-1}\right|\nn\\
&&\hspace{.3in}+\frac{1}{2}\log\left|\Ibf+\Hbf_4\widehat\Sbf_{2}^*\Hbf_4^T\left(\Ibf+\Hbf_3\widehat\Sbf_1^*\Hbf_3^T\right)^{-1}\right|\nn\\
&&\hspace{-.2in}\stackrel{(c)}\leq\frac{1}{2}\log\left|\Ibf+\Hbf_1\Sbf_{1}^*\Hbf_1^T\left(\Ibf+\Hbf_2\Sbf_2^*\Hbf_2^T\right)^{-1}\right|\nn\\
&&\hspace{.3in}+\frac{1}{2}\log\left|\Ibf+\Hbf_4\Sbf_{2}^*\Hbf_4^T\left(\Ibf+\Hbf_3\Sbf_1^*\Hbf_3^T\right)^{-1}\right|,\label{eq:Rs}\eqa
where in (a) we let \bqa
\Abf_1=\left(\Ibf+\Hbf_2\widehat\Sbf_2^*\Hbf_2^T\right)\Hbf_1^{-T}\Hbf_3^T,\label{eq:A1hat}\eqa
and
\bqa\Abf_2=\left(\Ibf+\Hbf_3\widehat\Sbf_1^*\Hbf_3^T\right)\Hbf_4^{-T}\Hbf_2^T.\label{eq:A2hat}\eqa
Equality (b) is from the fact
$\left|\Ibf-\Ubf\Vbf\right|=\left|\Ibf-\Vbf\Ubf\right|$.
Inequality (c) is from the assumption that $\Sbf_1^*$ and
$\Sbf_2^*$ optimize (\ref{eq:opt}) and the equality holds when
$\widehat\Sbf_1^*=\Sbf_1^*$ and $\widehat\Sbf_2^*=\Sbf_2^*$.

 The
above sum rate in (\ref{eq:Rs}) is also achievable by treating
interference as noise at each receiver, therefore the sum-rate
capacity is (\ref{eq:Rs}), if there exist Gaussian vectors $\np_1$
and $\np_2$ with distribution in (\ref{eq:jointZN}) that satisfies
(\ref{eq:equation1}), (\ref{eq:equation2}), (\ref{eq:A1hat}) and
(\ref{eq:A2hat}).

We consider the existence of $\np_1$. From Lemma \ref{lemma:pd},
$\np_1$ exists if and only if \bqa \Sigmabf_1\succeq
\Abf_1^T\Abf_1,\eqa with $\Abf_1$ defined in (\ref{eq:A1hat}).

From (\ref{eq:equation2}) and Woodbury identity
\cite{Golub&vanLoanV3:book}:\bqa
&&\hspace{-.4in}\left(\Abf+\Cbf\Bbf\Cbf^T\right)^{-1}\nn\\
&&\hspace{-.1in}=\Abf^{-1}-\Abf^{-1}\Cbf\left(\Bbf^{-1}+\Cbf^T\Abf^{-1}\Cbf\right)^{-1}\Cbf^T\Abf^{-1},\eqa
we have \bqa\hspace{-.0in}
\Sigmabf_2^{-1}=\Ibf-\Abf_1\left(-\Sigmabf_1+\Abf_1^T\Abf_1\right)^{-1}\Abf_1^T.\label{eq:sigma1inv}\eqa
On substituting (\ref{eq:sigma1inv}) into (\ref{eq:equation1}) we
have \bqa
&&\hspace{-.4in}\Sigmabf_1=\Ibf-\Abf_2\Abf_2^T+\Abf_2\Abf_1\left(\Abf_1^T\Abf_1-\Sigmabf_1\right)^{-1}\Abf_1^T\Abf_2^T.\label{eq:eqSigma1}\eqa
Define \bqa \Xbf=\Sigmabf_1-\Abf_1^T\Abf_1\eqa and substitute
(\ref{eq:M}) and (\ref{eq:W1}) into (\ref{eq:eqSigma1}). We then
have the following matrix equation: \bqa
\Xbf+\Wbf_1^T\Xbf^{-1}\Wbf_1=\Mbf.\label{eq:mxequation}\eqa
Equation (\ref{eq:mxequation}) is a special case of a discrete
algebraic Ricatti equation \cite{Engwerda-etal:93LA&A}. From Lemma
\ref{lemma:pdsolution}, with $\Mbf$ symmetric and positive
definite, (\ref{eq:mxequation}) has symmetric positive definite
solution $\Xbf$ if and only if (\ref{eq:condition1}) holds.
Therefore $\np_1$ exists with condition (\ref{eq:condition1}).
Similarly, $\np_2$ exists with condition (\ref{eq:condition2}).

Therefore if (\ref{eq:condition1}) and (\ref{eq:condition2}) hold
for any $\Sbf_i$ satisfying the power constraint, for any choice
of $\Sbf_{ij},i=1,2,j=1,\dots,n$, the sum rate must satisfy
(\ref{eq:Rs}). This completes our proof.

\subsection{Proof of Theorem \ref{theorem:NIsum2}}
In the proof of Theorem \ref{theorem:NIsum}, we let
(\ref{eq:equation1}) and (\ref{eq:equation2}) hold, and obtain
(\ref{eq:constdiff1}) and (\ref{eq:constdiff2}). On the other
hand, by Lemma \ref{lemma:opt}, if (\ref{eq:condition2_1}) and
(\ref{eq:condition2_2}) hold then we can still obtain
(\ref{eq:constdiff1}) and (\ref{eq:constdiff2}). The rest of the
proof of Theorem \ref{theorem:NIsum2} is the same as the proof of
Theorem \ref{theorem:NIsum}. Therefore, treating interference as
noise is sum-rate capacity achieving if there exist $\Sigmabf_1$
and $\Sigmabf_2$ that satisfy (\ref{eq:condition2_1}) and
(\ref{eq:condition2_2}).

\subsection{Proof of Theorem \ref{theorem:ZGIC}}

We provide two proofs of the first part of Theorem
\ref{theorem:ZGIC}, i.e., $\Hbf_2^T\Hbf_2\prec\Hbf_4^T\Hbf_4$. The
first proof  applies the same genie-aided method we used in the
proof of Theorem \ref{theorem:NIsum}.  The second proof does not
need a genie and is based on Lemma \ref{lemma:opt}.

\subsubsection{Genie-aided proof}
This proof is similar to the proof of Theorem \ref{theorem:NIsum}
but much simpler. Assume a Gaussian vector $\np$ which has joint
distribution with $\zp_1$ as \bqa \left[\begin{array}{c}
  \zp_1 \\
  \np \\
\end{array}\right]\sim\Nmat\left(\0bf,\left[\begin{array}{cc}
  \Ibf & \Abf \\
  \Abf^T & \Sigmabf \\
\end{array}\right]\right).\label{eq:covz}\eqa
Let $\np^n$ be a sequence of $n$ column random vectors with each
$\np_i$ being i.i.d. Then from Fano's inequality we have \bqa &&\hspace{-.2in}n(R_1+R_2)-n\epsilon\nn\\
&&\hspace{-.2in}\leq
I\left(\xp_1^n;\yp_1^n\right)+I\left(\xp_2^n;\yp_2^n\right)\nn\\
&&\hspace{-.2in}\leq
I\left(\xp_1^n;\yp_1^n,\Hbf_1\xp_1^n+\np^n\right)+I\left(\xp_2^n;\yp_2^n\right)\nn\\
&&\hspace{-.2in}=h\left(\Hbf_1\xp_1^n+\np^n\right)+h\left(\Hbf_1\xp_1^n+\Hbf_2\xp_2^n+\zp_1^n\left|\Hbf_1\xp_1^n+\np^n\right.\right)
\nn\\
&&\hspace{-.05in}-h\left(\Hbf_2\xp_2^n+\zp_1^n\left|\np^n\right.\right)+h\left(\Hbf_4\xp_2^n+\zp_2^n\right)-h\left(\np^n\right)-h\left(\zp_2^n\right)\nn\\
&&\hspace{-.2in}\stackrel{(a)}\leq nh\left(\Hbf_1\widehat\xp_{1}^*+\np\right)+nh\left(\Hbf_1\widehat\xp_{1}^*+n\Hbf_2\widehat\xp_{2}^*+\zp_1\left|\Hbf_1\widehat\xp_{1}^*+\np\right.\right)\nn\\
&&\hspace{-.05in}-h\left(\Hbf_2\xp_2^n+\zp_1^n\left|\np^n\right.\right)+h\left(\Hbf_4\xp_2^n+\zp_2^n\right)-nh\left(\np\right)-nh\left(\zp_2\right)\nn\\
&&\hspace{-.2in}\stackrel{(b)}\leq nh\left(\Hbf_1\widehat\xp_{1}^*+\np\right)+nh\left(\Hbf_1\widehat\xp_{1}^*+n\Hbf_2\widehat\xp_{2}^*+\zp_1\left|\Hbf_1\widehat\xp_{1}^*+\np\right.\right)\nn\\
&&\hspace{-.05in}-nh\left(\Hbf_2\widehat\xp_2^*+\zp_1\left|\np\right.\right)+nh\left(\Hbf_4\widehat\xp_2^*+\zp_2\right)-nh\left(\np\right)-nh\left(\zp_2\right)\nn\\
&&\hspace{-.2in}\stackrel{(c)}=\frac{n}{2}\left(\log\left|\Hbf_1\widehat\Sbf_1^*\Hbf_1^T+\Sigmabf\right|-\log\left|\Hbf_2\widehat\Sbf_2^*\Hbf_2^T+\Ibf-\Abf\Sigmabf^{-1}\Abf^T\right|\right.\nn\\
&&\hspace{-.05in}+\log\left|\Hbf_4\widehat\Sbf_2^*\Hbf_4^T+\Ibf\right|+\log\left|\Hbf_1\widehat\Sbf_1^*\Hbf_1^T+\Hbf_2\widehat\Sbf_2^*\Hbf_2^T+\Ibf\right.\nn\\
&&\hspace{-.05in}\left.-\left(\Hbf_1\widehat\Sbf_1^*\Hbf_1^T+\Abf\right)\left(\Hbf_1\widehat\Sbf_1^*\Hbf_1^T+\Sigmabf\right)^{-1}\left(\Hbf_1\widehat\Sbf_1^*\Hbf_1^T+\Abf^T\right)\right|\nn\\
&&\hspace{-.05in}\left.-\log\left|\Sigmabf\right|\right)\nn\\
&&\hspace{-.2in}=\frac{n}{2}\left(\log\left|\Ibf+\Hbf_1\widehat\Sbf_1^*\Hbf_1^T\Abf^{-1}\right|+\log\left|\Ibf+\Hbf_4\widehat\Sbf_2^*\Hbf_4^T\right|\right)\nn\\
&&\hspace{-.2in}=\frac{n}{2}\left(\log\left|\Ibf+\Hbf_1\widehat\Sbf_1^*\Hbf_1^T\left(\Ibf+\Hbf_2\widehat\Sbf_2^*\Hbf_2^T\right)^{-1}\right|\right.\nn\\
&&\hspace{1.8in}\left.+\log\left|\Ibf+\Hbf_4\widehat\Sbf_2^*\Hbf_4^T\right|\right)\nn\\
&&\hspace{-.2in}\leq\frac{n}{2}\cdot\max_{\substack{\textrm{tr}(\Sbf_1)\leq
P_1\\\textrm{tr}(\Sbf_2)\leq
P_2}}\left(\log\left|\Ibf+\Hbf_1\Sbf_1\Hbf_1^T\left(\Ibf+\Hbf_2\Sbf_2\Hbf_2^T\right)^{-1}\right|\right.\nn\\
&&\hspace{1.3in}\left.+\log\left|\Ibf+\Hbf_4\Sbf_2\Hbf_4^T\right|\right)\label{eq:RsZ}
\eqa
 where, (a) is from Lemmas \ref{lemma:noconditonal} and \ref{lemma:conditionaldirect}, and $\widehat\xp_i^*$ is zero mean Gaussian vector with
$\Cov\left(\xp^{*}_i\right)=\frac{1}{n}\sum_{j=1}^n\Cov\left(\xp_{i,j}\right)$,
$i=1,2$; in (b) we let \bqa
\Hbf_4^{-1}\Hbf_4^{-T}=\Hbf_2^{-1}\left(\Ibf-\Abf\Sigmabf^{-1}\Abf^T\right)\Hbf_2^{-T},\label{eq:cancelz}\eqa
and thus \bqa
&&-h\left(\Hbf_2\xp_2^n+\zp_1^n\left|\np\right.\right)+h\left(\Hbf_4\xp_2^n+\zp_2^n\right)\nn\\
&&=-n\log\left(abs\left|\Hbf_2\right|\right)+n\log\left(abs\left|\Hbf_4\right|\right)\nn\\
&&=-nh\left(\Hbf_2\widehat\xp_2^*+\zp_1\left|\np\right.\right)+nh\left(\Hbf_4\widehat\xp_2^*+\zp_2\right);\eqa
in (c) we let \bqa
\Abf=\Ibf+\Hbf_2\widehat\Sbf_2^*\Hbf_2.\label{eq:Az}\eqa

In order that all the equalities in (\ref{eq:RsZ}) hold, there
must exist $\np$ such that the covariance matrix in
(\ref{eq:covz}) satisfies (\ref{eq:cancelz}) and (\ref{eq:Az}).
From (\ref{eq:cancelz}) and (\ref{eq:Az}) we have \bqa
\Sigmabf=\Abf^T\left(\Ibf-\Hbf_2\Hbf_4^{-1}\Hbf_4^{-T}\Hbf_2^T\right)\Abf.\eqa
Therefore $\np$ exists if and only if \bqa
\Ibf-\Hbf_2\Hbf_4^{-1}\Hbf_4^{-T}\Hbf_2^T\succ\0bf,\eqa which is
equivalent to \bqa \Hbf_2^T\Hbf_2\prec\Hbf_4^T\Hbf_4.\eqa

\subsubsection{Proof based on Lemma \ref{lemma:opt}}
Starting from Fano's inequality we have \bqa &&n(R_1+R_2)-n\epsilon\nn\\
&&\leq
I\left(\xp_1^n;\yp_1^n\right)+I\left(\xp_2^n;\yp_2^n\right)\nn\\
&&=h\left(\Hbf_1\xp_1^n+\Hbf_2\xp_2^n+\zp_1^n\right)
-h\left(\Hbf_2\xp_2^n+\zp_1^n\right)\nn\\
&&\hspace{.3in}+h\left(\Hbf_4\xp_2^n+\zp_2^n\right)-h\left(\zp_2^n\right)\nn\\
&&\stackrel{(a)}\leq
nh\left(\Hbf_1\widehat\xp_1^{*}+\Hbf_2\widehat\xp_2^{*}+\zp_1\right)\nn\\
&&\hspace{.3in}-h\left(\xp_2^n+\Hbf_2^{-1}\zp_1^n\right)+h\left(\xp_2^n+\Hbf_4^{-1}\zp_2^n\right)-h\left(\zp_2^n\right)\nn\\
&&\hspace{.3in}-n\log\left(abs\left|\Hbf_2\right|\right)+n\log\left(abs\left|\Hbf_4\right|\right)\nn\\
 &&\stackrel{(b)}\leq nh\left(\Hbf_1\widehat\xp_1^{*}+\Hbf_2\widehat\xp_2^{*}+\zp_1^n\right)
-nh\left(\widehat\xp_2^*+\Hbf_2^{-1}\zp_1\right)\nn\\
&&\hspace{.3in}+nh\left(\widehat\xp_2^*+\Hbf_4^{-1}\zp_2\right)-h\left(\zp_2\right)\nn\\
&&\hspace{.3in}-n\log\left(abs\left|\Hbf_2\right|\right)+n\log\left(abs\left|\Hbf_4\right|\right)\nn\\
&&=\frac{n}{2}\log\left|\Ibf+\Hbf_1\widehat\Sbf_1^*\Hbf_1^T\left(\Ibf+\Hbf_2\widehat\Sbf_2\Hbf_2^T\right)^{-1}\right|\nn\\
&&\hspace{.3in}+\frac{n}{2}\log\left|\Ibf+\Hbf_4\widehat\Sbf_2^*\Hbf_4^T\right|\nn\\
&&\leq\max_{\substack{\textrm{tr}(\Sbf_1)\leq
P_1\\\textrm{tr}(\Sbf_2)\leq
P_2}}\left\{\frac{n}{2}\log\left|\Ibf+\Hbf_1\Sbf_1\Hbf_1^T\left(\Ibf+\Hbf_2\Sbf_2\Hbf_2^T\right)^{-1}\right|\right.\nn\\
&&\hspace{.8in}\left.+\frac{n}{2}\log\left|\Ibf+\Hbf_4\Sbf_2\Hbf_4^T\right|\right\},\label{eq:ZIC2}\eqa
where (a) is from Lemma \ref{lemma:noconditonal} and
$\widehat\xp_i^*$ is zero mean Gaussian vector with
$\Cov\left(\xp^{*}_i\right)=\frac{1}{n}\sum_{j=1}^n\Cov\left(\xp_{i,j}\right)$;
 (b) is from Lemma \ref{lemma:opt}.

 Next we prove the second part of Theorem \ref{theorem:ZGIC}. The
 achievability of the sum rate is straightforward by letting the
 first receiver decode both messages. We need only to show the
 converse. Start from Fano's inequality and notice that $\Hbf_2^{-1}\Hbf_2^{-T}\preceq\Hbf_4^{-1}\Hbf_4^{-T}$, then the second and third terms of (b) in
 (\ref{eq:ZIC2}) become \bqa
 &&-h\left(\xp_2^n+\Hbf_2^{-1}\zp_1^n\right)+h\left(\xp_2^n+\Hbf_4^{-1}\zp_2^n\right)\nn\\
 &&=-h\left(\xp_2^n+\Hbf_2^{-1}\zp_1^n\right)+h\left(\xp_2^n+\Hbf_2^{-1}\zp_1^n+\tilde\zp^n\right)\nn\\
 &&=I\left(\tilde\zp^n;\xp_2^n+\Hbf_2^{-1}\zp_1^n+\tilde\zp^n\right)\nn\\
 &&\leq I\left(\tilde\zp^n;\Hbf_2^{-1}\zp_1^n+\tilde\zp^n\right)\nn\\
 &&= -h\left(\Hbf_2^{-1}\zp_1^n\right)+h\left(\Hbf_4^{-1}\zp_2^n\right),\label{eq:ZIChalfDone}\eqa
 where
 $\tilde\zp\sim\Nmat\left(\0bf,\Hbf_4^{-1}\Hbf_4^{-T}-\Hbf_2^{-1}\Hbf_2^{-T}\right)$.
On substituting (\ref{eq:ZIChalfDone}) back into (\ref{eq:ZIC2})
we have \bqa
 R_1+R_2-\epsilon\leq\frac{1}{2}\log\left|\Ibf+\Hbf_1\widehat\Sbf_1^*\Hbf_1^T+\Hbf_2\widehat\Sbf_2\Hbf_2^T\right|.\label{eq:Zstrong1}
\eqa

On the other hand, we have\bqa  &&\hspace{-.2in}n(R_1+R_2)-n\epsilon\nn\\
&&\hspace{-.2in}\leq I\left(\xp_1^n;\yp_1^n\left|\xp_2^n\right.\right)+I\left(\xp_2^n;\yp_2^n\right)\nn\\
&&\hspace{-.2in}\leq\frac{1}{2}\log\left|\Ibf+\Hbf_1\widehat\Sbf_1^*\Hbf_1^T\right|+\frac{1}{2}\log\left|\Ibf+\Hbf_4\widehat\Sbf_2^*\Hbf_4^T\right|.\label{eq:Zstrong2}\eqa
From (\ref{eq:Zstrong1}) and (\ref{eq:Zstrong2}) we have \bqa
&&\hspace{-.3in}R_1+R_2-\epsilon\nn\\
&&\hspace{-.25in}\leq\min\left\{\begin{array}{c}
  \dfrac{1}{2}\log\left|\Ibf+\Hbf_1\widehat\Sbf_1\Hbf_1^T+\Hbf_2\widehat\Sbf_2\Hbf_2^T\right| \\
  \dfrac{1}{2}\log\left|\Ibf+\Hbf_1\widehat\Sbf_1\Hbf_1^T\right|+\dfrac{1}{2}\log\left|\Ibf+\Hbf_4\widehat\Sbf_2\Hbf_4^T\right| \\
\end{array}\right\}\nn\\
&&\hspace{-.25in}\leq\max_{\substack{\textrm{Cov}\left(\Sbf_1\right)\leq
P_1\\\textrm{Cov}\left(\Sbf_2\right)\leq
P_2}}\min\nn\\
&&\hspace{-.15in}\left\{\begin{array}{c}
  \dfrac{1}{2}\log\left|\Ibf+\Hbf_1\Sbf_1\Hbf_1^T+\Hbf_2\Sbf_2\Hbf_2^T\right| \\
  \dfrac{1}{2}\log\left|\Ibf+\Hbf_1\Sbf_1\Hbf_1^T\right|+\dfrac{1}{2}\log\left|\Ibf+\Hbf_4\Sbf_2\Hbf_4^T\right| \\
\end{array}\right\}.\eqa

\subsection{Proof of Theorem \ref{theorem:GICstrong}}

The achievability is straightforward by letting both receivers
decode both messages. We need only to show the converse, which can
be shown by setting $\Hbf_2=\0bf$ and $\Hbf_3=\0bf$, respectively,
and using Theorem \ref{theorem:ZGIC}.

\subsection{Proof of Theorem \ref{theorem:GICmixed}}

The achievability part is straightforward by letting user $2$
first decode message from user $1$ and then decode its own
message, and user $1$ treat signals from user $2$ as noise.

To prove the converse, we first let $\Hbf_3=\0bf$ and use the
first part of Theorem \ref{theorem:ZGIC}, and then let
$\Hbf_2=\0bf$ and use the second part of Theorem
\ref{theorem:ZGIC}. We obtain \bqa
&&\hspace{-.3in}R_1+R_2\leq\max_{\textrm{tr}\left(\Sbf_1\right)\leq
P_1,\textrm{tr}\left(\Sbf_2\right)\leq
P_2}\min\nn\\
&&\hspace{-.5in}\left\{\begin{array}{c}
\dfrac{1}{2}\log\left|\Ibf+\Hbf_3\Sbf_1\Hbf_3^T+\Hbf_4\Sbf_2\Hbf_4^T\right|, \\
\hspace{-.4in}\dfrac{1}{2}\log\left|\Ibf+\Hbf_1\Sbf_1\Hbf_1^T\left(\Ibf+\Hbf_2\Sbf_2\Hbf_2^T\right)^{-1}\right|\\
\hspace{1.2in}+\dfrac{1}{2}\log\left|\Ibf+\Hbf_4\Sbf_2\Hbf_4^T\right|,\\
  \dfrac{1}{2}\log\left|\Ibf+\Hbf_1\Sbf_1\Hbf_1^T\right|+\dfrac{1}{2}\log\left|\Ibf+\Hbf_4\Sbf_2\Hbf_4^T\right| \\
  \end{array}\right\}.\label{eq:GICmixed}\eqa
We complete the proof by pointing out that the last line of
(\ref{eq:GICmixed}) is redundant because of the second line.

\subsection{Proof of Theorems \ref{theorem:ZICstrongRegion},
\ref{theorem:ICstrongRegion} and \ref{theorem:PGIC}}

Theorems \ref{theorem:ZICstrongRegion} and
\ref{theorem:ICstrongRegion} are consequences of Theorems
\ref{theorem:ZGIC} and \ref{theorem:GICstrong} respectively.  The
proof is straightforward and hence is omitted.

The proof of Theorem \ref{theorem:PGIC} is also omitted due to the
lack of space.

\section{Numerical Result}

\begin{figure}[htp]
\vspace{-.1in}\centerline{\leavevmode \epsfxsize=3.6in
\epsfysize=3in
\epsfbox{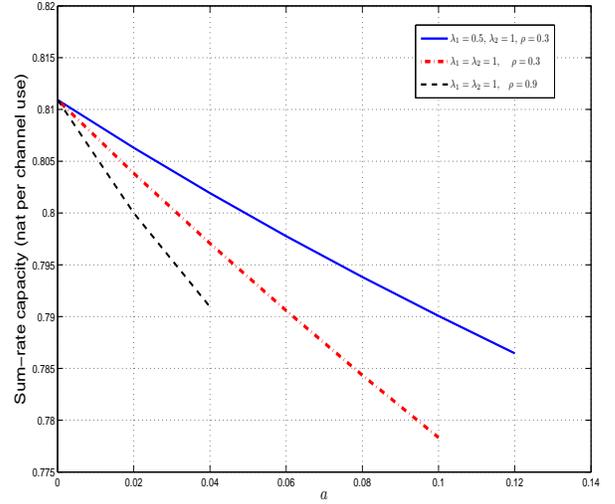}}\vspace{-.2in}\caption{Sum-rate capacity
v.s. $a$} \label{fig:sum}\end{figure}

Consider a symmetric MIMO IC with two transmit antennas and two
receive antennas. Let $\Hbf_1=\Hbf_2=\Ibf$,
$\Hbf_2=\Hbf_3=\sqrt{a}\left[\begin{array}{cc}
  \lambda_1 & \rho \\
  \rho & \lambda_2 \\
\end{array}\right]$, where $a$ varies from $0$ to $1$. Fig.
\ref{fig:sum} shows the noisy-interference sum-rate capacity v.s.
$a$, for different $\lambda_1,\lambda_2$ and $\rho$. There is a
range of $a$, within which the channel has noisy interference.
Fig. \ref{fig:sum} shows that the range of $a$ and the sum-rate
capacity decrease as the norm of $\Hbf_2$ and $\Hbf_3$ increases.

\section{Conclusion}

We have extended the capacity results on scalar ICs to MIMO ICs
and have obtained the sum-rate capacity of the MIMO IC with
noisy-interference, strong interference, and Z-interference, and
the capacity region of the MIMO IC with strong interference.

\section*{Acknowledgment}


We gratefully acknowledge S. Annapureddy and V. Veeravalli from
the University of Illinois at Urbana Champaign for pointing out an
error in the original manuscript related to Theorems
\ref{theorem:NIsum} and \ref{theorem:NIsum2}.

\section*{Appendix}

\subsection{Proof of Lemma \ref{lemma:generalconcave}}
Define a discrete random variable $E$ with the distribution \bqa
P_E(E=i)=\lambda_i,\quad i=1,\dots,k.\eqa Let the conditional
distribution of $\xp^n$ be \bqa
p_{\xp^n|E}\left(\xp^n\left|E=i\right.\right)=
  p_{\xp^n_i}\left(\xp^n_i\right).
\label{eq:ZgivenE}\eqa Then the probability density function of
$\xp^n$ is \bqa p_{\xp^n}=\sum_{i=1}^k\left(\lambda_i\cdot
p_{\xp^n_i}\right).\eqa Therefore \bqa
\Cov\left(\xp^{n}\right)=\sum_{i=1}^k\lambda_i\Cov\left(\xp_i^n\right)=\Cov\left(\yp^n\right).\eqa
Then from Lemma \ref{lemma:conditional2} we have \bqa
h\left(\xp_{\Smat}\left|\xp_{\bar\Smat}\right.\right)\leq
h\left(\yp_{\Smat}\left|\yp_{\bar\Smat}\right.\right).\eqa From
(\ref{eq:ZgivenE}) we have \bqa
&&h\left(\xp_{\Smat}\left|\xp_{\bar\Smat},E\right.\right)\nn\\
&&=\sum_{i=1}^kP_E(E=i)
h\left(\xp_{i,\Smat}\left|\xp_{i,\bar\Smat}\right.\right)\nn\\
&&=\sum_{i=1}^k\lambda_i
h\left(\xp_{i,\Smat}\left|\xp_{i,\bar\Smat}\right.\right)\nn\\
&&\leq h\left(\xp_{\Smat}\left|\xp_{\bar\Smat}\right.\right). \eqa
Therefore we have \bqa \sum_{i=1}^k\lambda_i
h\left(\xp_{i,\Smat}\left|\xp_{i,\bar\Smat}\right.\right)\leq
h\left(\yp_{\Smat}\left|\yp_{\bar\Smat}\right.\right).\eqa

\subsection{Proof of Lemma \ref{lemma:opt}}\vspace{-.3in}
\bqa
&&h\left(\xp^n+\zp^n\right)-h\left(\xp^n+\zp^n+\tilde\zp^n\right)\nn\\
&&=-I\left(\tilde\zp^n;\xp^n+\zp^n+\tilde\zp^n\right)\nn\\
&&\stackrel{(a)}\leq-I\left(\tilde\zp^n;\xp^{*n}+\zp^n+\tilde\zp^n\right)\nn\\
&&=-h\left(\tilde\zp^n\right)+h\left(\tilde\zp^n\left|\xp^{*n}+\zp^n+\tilde\zp^n\right.\right)\nn\\
&&\stackrel{(b)}\leq
-nh\left(\tilde\zp\right)+nh\left(\tilde\zp\left|\widehat\xp^*+\zp+\tilde\zp\right.\right)\nn\\
&&=-nI\left(\tilde\zp;\widehat\xp^*+\zp+\tilde\zp\right)\nn\\
&&=nh\left(\widehat\xp^*+\zp\right)-nh\left(\widehat\xp^*+\zp+\tilde\zp\right),\eqa
where (a) is from Lemma \ref{lemma:worstnoise} and $\xp^{*n}$ has
the same covariance matrix as $\xp^{n}$, and (b) is from Lemma
\ref{lemma:conditionaldirect}.

\bibliography{Journal,Conf,Misc,Book}
\bibliographystyle{IEEEbib}
\end{document}

%% file: macro.tex
\newtheorem{theorem}{Theorem}
\newtheorem{example}{Example}

\newtheorem{lemma}{Lemma}
\newtheorem{definition}{Definition}

\def\psfancypar#1#2{\begingroup\def\par{\endgraf\endgroup\lineskiplimit=0pt}
               \setbox2=\hbox{\large\sc #2}
               \newdimen\tmpht \tmpht \ht2 \advance\tmpht by \baselineskip
               \font\hhuge=Times-Bold at \tmpht
               \setbox1=\hbox{{\hhuge #1}}
               \count7=\tmpht \count8=\ht1
               \divide\count8 by 1000 \divide\count7 by \count8
               \tmpht=.001\tmpht\multiply\tmpht by \count7
               \font\hhuge=Times-Bold at \tmpht
               \setbox1=\hbox{{\hhuge #1}}
               \noindent
                \hangindent1.05\wd1
               \hangafter=-2 {\hskip-\hangindent
               \lower1\ht1\hbox{\raise1.0\ht2\copy1}%
                \kern-0\wd1}\copy2\lineskiplimit=-1000pt}

\newenvironment{ap}{
\renewcommand{\thesection}{\mbox{Appendix }\Alph{section}}
\renewcommand{\thesubsection}{\Alph{section}.\arabic{subsection
}}
\renewcommand{\theequation}{\mbox{\Alph{section}.}\arabic{equation}}
}{
\renewcommand{\thesection}{\arabic{section}}
\renewcommand{\thesubsection}{\thesection.\arabic{subsection}}
\renewcommand{\theequation}{\arabic{equation}}
}

\newcommand{\beq}{\begin{equation}}
\newcommand{\eeq}{\end{equation}}
\newcommand{\bqa}{\begin{eqnarray}}
\newcommand{\eqa}{\end{eqnarray}}
\newcommand{\bqn}{\begin{eqnarray*}}
\newcommand{\eqn}{\end{eqnarray*}}
\newcommand{\nn}{\nonumber}

\newcommand{\be}{\begin{enumerate}}
\newcommand{\ee}{\end{enumerate}}
\newcommand{\bi}{\begin{itemize}}
\newcommand{\ei}{\end{itemize}}
\newcommand{\bd}{\begin{description}}
\newcommand{\ed}{\end{description}}
\newcommand{\ba}{\begin{array}}
\newcommand{\ea}{\end{array}}
\newcommand{\bde}{\begin{definition}}
\newcommand{\ede}{\end{definition}}
\newcommand{\bex}{\begin{example}}
\newcommand{\eex}{\end{example}}

\def\boxit#1{\vbox{\hrule\hbox{\vrule\kern3pt
        \vbox{\kern3pt#1\kern3pt}\kern3pt\vrule}\hrule}}

\def\reals{ { {\rm  I \kern-0.15em R }  } }
\def\complex{ {\,{{\rm C} \kern-0.50em \raise0.20ex {  |}}\, }}
\def\alphabf{\hbox{\boldmath$\alpha$\unboldmath}}

\def\Sigmabf{\hbox{$\bf \Sigma$}}

\def\Sigmabf{\mbox{$ \bf \Sigma $}}

\def\0bf{{\bf 0}}
\def\1bf{{\bf 1}}
\def\2bf{{\bf 2}}
\def\3bf{{\bf 3}}
\def\4bf{{\bf 4}}
\def\5bf{{\bf 5}}
\def\6bf{{\bf 6}}
\def\7bf{{\bf 7}}
\def\8bf{{\bf 8}}
\def\9bf{{\bf 9}}

\def\Abf{{\bf A}}
\def\Bbf{{\bf B}}
\def\Cbf{{\bf C}}

\def\Hbf{{\bf H}}
\def\Ibf{{\bf I}}

\def\Kbf{{\bf K}}

\def\Mbf{{\bf M}}

\def\Rbf{{\bf R}}
\def\Sbf{{\bf S}}

\def\Ubf{{\bf U}}
\def\Vbf{{\bf V}}
\def\Wbf{{\bf W}}
\def\Xbf{{\bf X}}

\def\np{{\pmb n}}

\def\xp{{\pmb x}}
\def\yp{{\pmb y}}
\def\zp{{\pmb z}}

\def\Nmat{\mathcal{N}}

\def\Smat{\mathcal{S}}

\def\QED{\mbox{\rule[0pt]{1.3ex}{1.3ex}}}

\def\Rxx{\Rbf_{\ssstyle X\kern-.1em X}}

\let\ssstyle=\scriptscriptstyle

\def\etal{{\it et al. \/}}

\def\Cov{{\textrm{Cov}}}
\def\tr{{\textrm{tr}}}

\def\Kout{\setbox1=\hbox{\Huge\bf K}\hbox to
1.05\wd1{\hspace{.05\wd1}
}}
\def\Sout{\setbox1=\hbox{\Huge\bf S}\hbox to 1.05\wd1{\hspace{.05\wd1}
}}